\documentclass[
    ,final            % use final for the camera ready runs
  ]
  {aipproc}

\layoutstyle{6x9}

\begin{document}

\title{The Final Merger of Comparable Mass Binary Black Holes}


\author{Joan M. Centrella}{
  address={Lab. For Gravitational Astrophysics, NASA Goddard
Space Flight Center, Greenbelt, MD 20771}
}

\begin{abstract}
 A remarkable series of breakthroughs in numerical relativity
modeling of black hole binary mergers has occurred over the
past few years.  This paper provides a general overview of
these exciting developments, focusing on recent progress in
merger simulations and calculations of the resulting gravitational
waveforms.
\end{abstract}

\maketitle

%%%%%%%%%%%%%%%%%%%%%%%%%%%%%%%%%%%%%%%%%%%%
%% MAINMATTER
%%%%%%%%%%%%%%%%%%%%%%%%%%%%%%%%%%%%%%%%%%%%

\section{Introduction}

The coalescence of a comparable mass black hole binary (BHB)
is a powerful source of gravitational waves and proceeds in 
three phases: inspiral, merger, and ringdown \cite{Flanagan97a}.  
The inspiral stage, when the BHs are widely separated and
following quasi-circular trajectories, and the 
ringdown stage, during which the final merged BH settles
into a quiescent Kerr state, can both be treated analytically.
However the merger phase, in which the two BHs plunge together
and merge to form a highly distorted remnant BH, occurs in the
arena of very strong, dynamical gravitational fields and can
only be calculated using numerical relativity.

This final merger will produce an intense burst of gravitational
radiation with a luminosity $\sim 10^{23} L_{\odot}$, briefly
emitting more energy than the combined light from all the stars
in the visible universe.  Such bursts are expected to be among
the strongest sources for LISA, which will observe mergers of massive
BHBs.  Mergers of stellar mass and intermediate mass BHBs
are likely to be the strongest sources for ground-based gravitational
wave detectors such as LIGO and VIRGO.  Observing the gravitational
waves from the final merger will allow unprecedented tests of
general relativity in the dynamical, strong field regime -- 
provided we know the waveforms that general relativity predicts.

The final merger of BHBs also has compelling astrophysical
implications.  In particular, when the BHs have unequal masses,
the resulting gravitational wave emission is asymmetric; since
the gravitational waves carry momentum, the merged remnant BH
suffers a recoil kick \cite{Favata:2004wz}.  
If this kick is large enough, it could
eject the merged remnant from its host structure, thereby
affecting the overall rate of merger 
events \cite{Merritt:2004xa}.  In addition,
since BHBs are generally expected to be spinning, their mergers
could produce interesting spin dynamics and couplings \cite{Merritt:2002hc}.

For more than 30 years, numerical relativists have attempted to calculate
the merger of comparable mass BHs and the resulting gravitational
waveforms.  This has proved to be a very difficult undertaking indeed.
In particular, the simulation codes have been plagued by a host
of difficulties, typically resulting in various instabilities that
caused them to crash before any sizeable fraction of a binary 
orbit could be evolved.

Recently, however, a series of dramatic breakthroughs has
occurred in numerical relativity, resulting in accurate and robust
simulations of BHB mergers and their gravitational waveforms.
This paper provides a general overview of these exciting developments.
We begin with a brief overview of numerical relativity and
BHB calculations.  The heart of the paper then focuses on recent progress
in computing BHB orbits and mergers, and the resulting waveforms.
The paper concludes with a summary and outlook for the future.

With the goal of reaching as wide an audience as possible, 
technical details are deliberately kept to a minimum; interested
readers can find more detailed information in the references.\footnote{Since
this paper is not a full review of the subject, the reference list
is representative rather than comprehensive.  We attempted to cite
key papers from the major numerical relativity efforts that
were steppingstones to the current breakthroughs.}
We follow conventional practice by setting $G = 1$ and $c = 1$,
which allows us to measure both time and distance in terms of
mass $M$.  In particular, $1 M \sim (5 \times 10^{-6})(M/M_{\odot}){\rm sec}
\sim 1.5 (M/M_{\odot})  {\rm km}$.  
Spatial indices are taken to have the range $i = 1,2,3$.
Note that the simulation results scale
with the masses of the BHs, and are thus applicable to
LISA as well as to ground-based detectors.

\section{Numerical Relativity}

Numerical relativists construct a spacetime by solving the
Einstein equations on a computer. 
In the most commonly used ``3+1'' \cite{Arnowitt62,Misner73}
 approach, 4-D spacetime is considered
to be sliced into a stack of 
 3-D spacelike hypersurfaces labeled by time $t$.
The main independent variables are essentially the 3-metric
$g_{ij}$  and its first time derivative
$ \partial_t g_{ij}$ on each slice.
The equations split naturally into
two sets. The constraint equations provide relationships that must
be satisfied at any time $t$; in particular,
initial data
for BHBs is set by solving the constraint equations on a 3-D slice at
some initial time $t=0$.  The evolution
equations are used to propagate this data forward in time.  The four coordinate
degrees of freedom in general relativity give four freely-specifiable
coordinate or gauge conditions for the 
future development of the time and
spatial coordinates. During the evolution
the gauge is specified by the lapse function
$\alpha$, which gives the lapse of proper time $\alpha \Delta t$ between
neighboring slices, and the shift vector $\beta^i$,
which governs
how the spatial coordinates develop from one slice to the next.

Efforts to evolve the merger of two BHs have a long 
history.\footnote{An excellent history of the developments in numerical
relativity treatments of the BHB problem can be found in
the talk by Miguel Alcubierre at the Astrophysical Applications
of Numerical Relativity workshop held in May 2006 in 
Guanajuato, Mexico.}
The first attempt to solve the Einstein equations on a computer was
carried out
 by Hahn and Lindquist in 1964 \cite{Hahn64}, who tried to evolve the
head-on collision of two equal mass BHs.  (Since the term ``black
hole'' had not yet been coined, they called their paper ``The two-body
problem in geometrodynamics.'')  Due in part to a poor choice of
coordinate conditions, the evolution crashed shortly after it began.
In the mid-1970s, Smarr and Eppley~\cite{Smarr76,Smarr77,Smarr79}
 pioneered the use of the
 the 3+1 approach with improved
coordinate conditions, including conditions on the lapse function
to produce slices that avoid crashing into 
singularities \cite{Smarr78b,Smarr78a}.  
Although their simulations encountered
instabilities and had problems with accuracy, they 
were able to evolve the head-on collision and
extract some information about the resulting
gravitational waves. Following this significant achievement there was
very little work on BHB simulations throughout the 1980s, 
although some numerical relativity
work did continue, mostly on neutron stars.  

In the 1990s work on ground-based gravitational wave detectors 
such as LIGO moved
ahead strongly; since BHB mergers are 
considered one of the most promising sources
for these detectors, numerical relativity work on
the BHB problem started up again.  
More accurate calculations of head-on collisions
were carried out, starting in axisymmetry~\cite{Anninos94b}.
In the mid-1990s,
the NSF funded the Binary Black Hole
 Grand Challenge collaboration, a large 
multi-institution effort aimed at evolving 
 BHBs in 3-D and calculating the resulting gravitational wave 
signatures. A vigorous numerical relativity program was also
started at the newly-formed Albert Einstein Insitutue
in Germany. While many important developments resulted from this
era, including the development of
 large 3-D codes and the ability to evolve boosted
BHs~\cite{Cook97a} and grazing 
collisions~\cite{Bruegmann97,Brandt00,Alcubierre00b}, 
the problem turned out to be more
difficult than anticipated and the codes were plagued by 
instabilities that caused them to crash.

During the late 1990s and early 2000s, the ground-based detectors
began taking data, 
the importance of BHB mergers as sources for LISA grew, and
new research groups arose in numerical relativity. The role
of unstable modes present in the formulations of the numerical
relativity equations was recognized as a major issue. Work on
key areas such as gauge
conditions, formalisms, boundary conditions, and the role of the
constraints in evolutions was carried out.
Overall, progress in obtaining stable 3-D BH evolutions was 
slow and incremental.
 The Lazarus approach  combined a brief 3-D
numerical relativity 
evolution of a BHB near the final plunge with a late-time
perturbative evolution to make use of the short-duration
stable evolutions that were then possible, and produced the first
gravitational waveform from a BHB~\cite{Baker:2001nu,Baker:2002qf}.

Most of the recent work in numerical relativity 
has been carried out using a conformal formulation of the Einstein
equations known as BSSN~\cite{Shibata95,Baumgarte99}.  
In this approach, the set of
evolution equations has first-order time derivatives and
second-order spatial derivatives, and is strongly hyperbolic
\cite{Nagy:2004td,Reula:2004xd}.
The constraint equations have been incorporated
into the evolution equations to improve the performance. In some
cases the BHs are represented as ``punctures,'' with the singular
parts being factored out \cite{Brandt97b}; 
in other cases, the BH interiors
are excised to remove the singular parts from the 
grid \cite{Shoemaker:2003td,Alcubierre:2004bm}.  Various
gauge conditions were developed to allow longer evolutions,
including new slicing conditions that avoid evolving into a
singularity and shift conditions that prevent the coordinates
from falling into the BHs \cite{Alcubierre02a}.  Other approaches feature fully
first-order symmetric hyperbolic formulations of the Einstein
equations and special attention to constraint preserving
boundary conditions \cite{Lindblom:2004gd}.

For successful BHB merger simulations, it is
necessary to resolve both the BHs (with spatial scales 
$\sim M$) and extract the gravitational radiation (with 
scales $\lambda_{\rm GW} \sim (10 - 100)M$) in the wave zone.
Since the large 3-D codes strain the capacities of current
high performance computing facilities, this requires the use of
variable resolution within the computational domain.  Most of
the current numerical relativity codes use finite differences on
a 3-D Cartesian grid with fixed or adaptive mesh refinement. 
There are also a few efforts that use spectral methods, which
also incorporate variable resolution.

In the past two years,
there has been significant and rapid progress 
in numerical relativity simulations of BHB mergers across a broad
front.
The first complete  orbit of a BHB was
achieved in 2004.  This was followed shortly by the first
simulation of a BHB through an orbit, plunge, merger and ringdown.
Since late 2005, new ideas have opened the field up even more, and
the past year has seen dramatic progress.  These developments
are reviewed in the next section.  

\section{BHB Orbits and Mergers}

The first complete orbit of an equal mass, nonspinning BHB binary
was achieved by Br\"ugmann, Tichy, and Jansen~\cite{Bruegmann:2003aw} using
the standard conformal BSSN approach with 
the BHs represented as ``punctures''~\cite{Brandt97b}.
  The 3-metric on the initial
slice is written as $g_{ij} = \psi^4 \delta_{ij}$, where the conformal
factor $\psi = \psi_{\rm BL} + u$ and $i,j = 1,2,3$.  
The static, 
singlar part of the conformal factor has the form 
$\psi_{\rm BL} = 1 + \sum_{n=1}^{2} m_n/2 |\vec{r} - \vec{r}_n|$,
where the $n^{\rm th}$  puncture black hole has mass $m_n$ and
is located at $\vec{r}_n$. The nonsingular function $u$ is obtained
by solving one of the constraint equations.
    
In the standard puncture approach, the
singular part of the metic is factored out and handled analytically
during the evolution,
and only the regular parts are evolved numerically.
This requires that the punctures remain fixed on the numerical
grid, resulting in a stretching of the
coordinate system and the development of large errors in the metric
as the binary evolves. Excision of the regions around the punctures
(but within the horizons) can be used to reduce errors and prolong
these runs; excision can be applied to the individual punctures
as well as inside a common horizon at late times.

For head-on and grazing collisions starting from relatively close
separations, these methods work fairly well, as a common horizon
forms quickly and excision prevents the unbounded growth of errors
and allows the simulations to continue long enough for the BHs to merge.
For orbiting BHs, a corotating coordinate frame implemented by
an angular shift vector is needed since the punctures remain fixed on
the grid.  This can cause serious problems, however, such as
superluminal coordinate speeds at large distances from the BHs
and incoming noise from the outer boundary of Cartesian grids.

Br\"ugmann, Tichy, and Jansen~\cite{Bruegmann:2003aw} introduced
comoving coordinates using a shift vector that is dynamically 
adjusted during the evolution of the BHB to minimize both the angular 
and radial motion of the BHs.  Their code uses fixed mesh refinement
implemented by nested Cartesian boxes, with the resolution decreasing
for boxes that span successively larger regions of the domain.
They were able to evolve a BHB using excised punctures for 
more than one orbit, to $t \sim 
185M$, where $M$ is the total mass of the system.
The code did crash before the BHs merged and inaccuracies
in the outer regions prevented the extraction of
gravitational waves. Their paper first appeared as a preprint
in December 2003.
Later, more accurate work by Diener, et al.\cite{Diener:2005mg}
highlights the importance of high resolution and the effects of
gauge choices on the resulting evolutions.
 Nevertheless, this first simulation of a full BHB orbit 
was a major step forward.

In the first part of 2005, Pretorius carried out the first evolution
of a BHB through a single plunge orbit, merger and ringdown
\cite{Pretorius:2005gq} using an approach completely different
from the standard one \cite{Pretorius:2006tp,Pretorius:2004jg}.  
Instead of using the 3+1 technique,
he evolves the 4-metric directly, using generalized harmonic
coordinates.  The evolution equations have second-order time derivatives
and constraint damping terms designed to remove spurious
non-physical solutions.
Numerical dissipation is added to control high frequency instabilities.
 
His initial data consists of two Lorentz-boosted scalar field
profiles, with positions and velocities
 chosen to approximate a BHB orbit.  Each
scalar field configuration quickly collapses to form a BH, yielding
a BHB system.  The BH interiors are excised and the BHs move
freely across the grid as the binary evolves; no corotating 
coordinates are needed.  Adaptive mesh refinement is used to
provide higher resolution in the regions near the BHs.  Outside
the orbital region, fixed mesh refinement is implemented using
nested Cartesian boxes centered on the origin.

\begin{figure}[t]
 \includegraphics[width=8.2cm,clip=true]{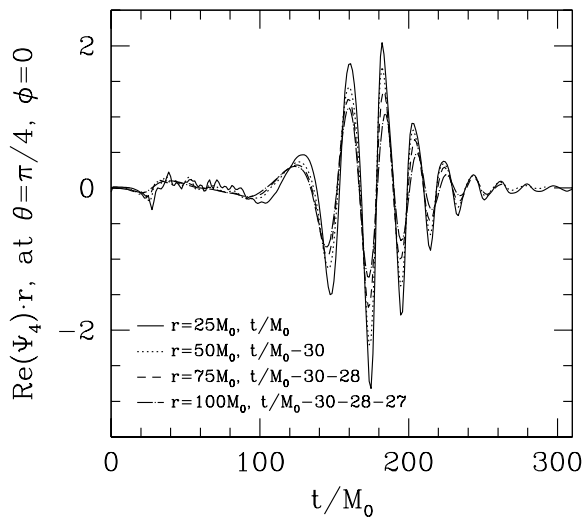}
  \caption{Waveforms from Pretorius' simulation \cite{Pretorius:2005gq}.}
\label{fig:Pretorius}
\end{figure}

\begin{figure}
 \includegraphics[height=.36\textheight]{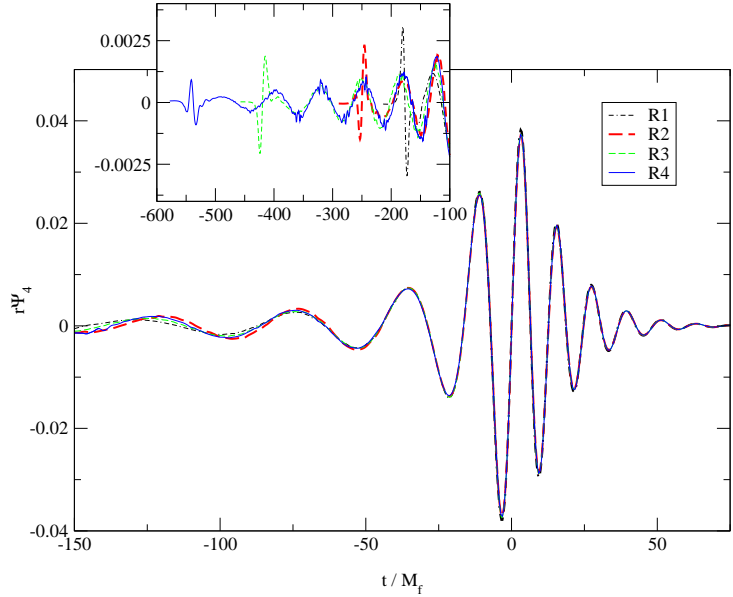}
  \caption{Waveforms from the long duration runs 
by the Goddard group, with parameters given in Table~\ref{tab:a}
\cite{Baker:2006yw}}.
\label{fig:waves}
\end{figure}

In Pretorius' simulation the BHs plunge together, completing
$\sim 1$ orbit before merging.  The code continues to run stably
during the subsequent ringdown of the merged remnant BH, allowing
the emitted gravitational waves to propagate outward far enough
to be extracted. Figure~\ref{fig:Pretorius} is taken from 
Ref.~\cite{Pretorius:2005gq} and shows the gravitational 
waveforms from this simulation as represented by $r\Psi_4$,
where $\Psi_4$ is the real part of the Weyl tensor component
and $r$ is the coordinate distance from the center of the
grid.\footnote{For comparison with the waveforms shown in 
Figure~\ref{fig:waves}, note that Pretorius uses the time coordinate
$t/M_0$, where $M_0$ is the mass of a single BH.  Also, the amplitude
of $r\Psi_4$ has not been scaled with $M$; when this is done, 
the amplitude is comparable with that shown in Figure~\ref{fig:waves}.}

This achievement set a new standard in numerical
relativity.  And the use of these nonstandard techniques raised
the important question of whether such methods were essential
for successful BHB merger simulations.

In late 2005 this question was answered when the numerical
relativity groups from the University of Texas at Brownsville
(UTB) \cite{Campanelli:2005dd}
 and NASA's Goddard Space Flight Center \cite{Baker:2005vv} simultaneously
and independently discovered a way to evolve a BHB through
an orbit, plunge, and merger within
the 3 + 1 approach.  Both of their codes 
are based on the BSSN formalism, with
a key difference.  Initially the BHs are set up using the
standard puncture technique.  However, during the evolution
the singular part of the conformal factor is {\em not} factored out
but rather is evolved together with the nonsingular part,
using regularization to handle the puncture singularities.
This allows the puncture BHs to move freely across the grid;
no excision or corotating coordinates
are needed.

Important ingredients in the
success of the moving puncture method are the  novel coordinate
conditions for the lapse and shift.  The UTB and Goddard groups
each developed somewhat different lapse and shift conditions;
both codes produced stable and accurate evolutions of a BHB
through the final plunge, merger, and ringdown.  The UTB code
used a particular coodinate transformation to allow the grid
resolution to vary smoothly over the computational domain,
whereas the Goddard code employed fixed mesh refinement based
on nested Cartesian boxes.  Both groups were able to extract
accurate and convergent gravitational waveforms.

The first results from moving puncture evolutions were presented 
by the UTB and Goddard groups in early November 2005 at the
``Numerical Relativity 2005'' 
workshop,\footnote{The presentations from this meeting
are posted at http://astrogravs.gsfc.nasa.gov/conf/numrel2005/.}
and the papers were submitted shortly thereafter.  Since these
techniques were developed within the widely-used traditional
3+1 numerical relativity approach, they could be readily adopted
by other groups with similar 3-D BSSN codes.  Indeed, in early January
2006 the Penn State group submitted a paper applying these techniques
to evolutions of nonequal mass BHBs \cite{Herrmann:2006ks}.  
The UTB and Goddard groups 
moved quickly and were soon able to evolve
multiple orbits followed by merger and ringdown 
\cite{Campanelli:2006gf,Baker:2006yw,vanMeter:2006vi}.
At the April 2006 APS meeting, an entire session was devoted to 
BHB merger simulations with moving punctures. More groups adopted
the moving puncture technique and, by the summer of 2006,
this method was being actively used, studied, and advanced by
the majority of the numerical relativity community working
on BHB simulations.

\begin{figure}[t]
 \includegraphics[height=.36\textheight]{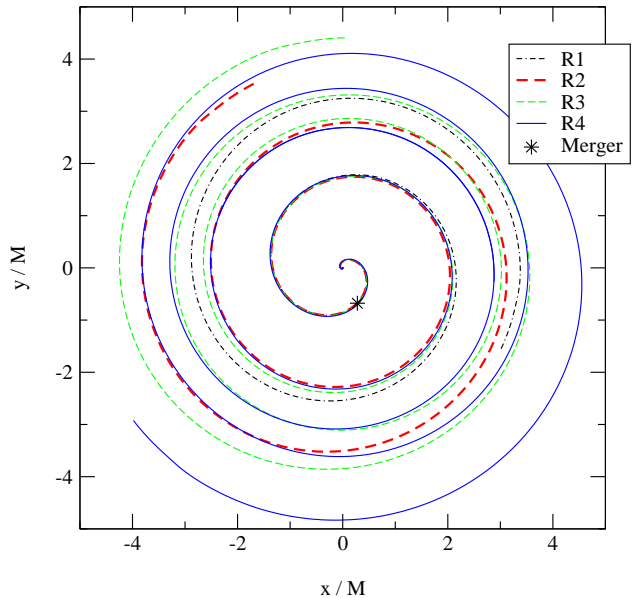}
  \caption{Paths of BHs for the long duration runs by the Goddard
group, with parameters given in Table~\ref{tab:a}.
For clarity, only the track of one BH from each run is 
shown \cite{Baker:2006yw}.
}
\label{fig:trajectories}
\end{figure}

In the early spring of 2006, the Goddard group used the moving
puncture method to study the dynamics and radiation generation 
in the last few orbits and merger of an equal mass nonspinning
BHB \cite{Baker:2006yw}.  Using adaptive mesh refinement
 to follow the BHs in the orbital region and fixed
mesh refinement based on nested Cartesian boxes in the outer regions,
they ran a series of long duration runs starting from successively
wider separations.  In each case, the BHs start out on approximately
circular orbits; key parameters for these runs are given in
Table~\ref{tab:a}.  Here, $L/M_0$ is the initial proper separation
of the BHs,\footnote{For the Goddard runs, $M_0$ is the initial total
system mass.} 
$T_{\rm sim}$ is the duration of each run, and
$T_{\rm merger}$ is the time at which the merger
occurs, starting from the initial time in each
run.  The number of orbits $N_{\rm orbits}$ is
estimated from the trajectories;
see Figure~\ref{fig:trajectories}.

\begin{table}[t]
\begin{tabular}{lrrrr}
\hline
   \tablehead{1}{r}{b}{Run}
  & \tablehead{1}{r}{b}{$L/M_0$}
  & \tablehead{1}{r}{b}{$T_{\rm sim}$}
  & \tablehead{1}{r}{b}{$T_{\rm merger}$}
  & \tablehead{1}{r}{b}{$N_{\rm orbits}$}   \\
\hline
R1 & 9.9 & $421M$ & $160M$  & 1.8\\
R2 & 11.1 & $531M$ & $234M$  & 2.5\\
R3 & 12.1 & $530M$ & $396M$ & 3.6\\
R4 & 13.2 & $850M$ & $513M$ & 4.2\\
\hline
\end{tabular}
\caption{Parameters of long duration BHB runs calculated by the 
Goddard group using the moving puncture method \cite{Baker:2006yw}.
}
\label{tab:a}
\end{table}

Note that these initial data sets are an approximation to the
actual conditions of a BHB in the real universe. Astrophysically,
a BHB quickly circularizes and then spirals together, emitting
gravitational waves, through $> 10^3$ nearly circular orbits
before the final plunge and merger.  Ideally, one would start a
numerical relativity simulation of a BHB with initial data that uses
BH positions and velocities from an 
inspiralling astrophysical orbit, and includes
the outgoing gravitational radiation from earlier parts of
the inspiral.  If this were possible, one could carry out a sequence
of simulations, starting the BHs with successively wider separations,
and expect the BHs to follow the same astrophysical trajectories
all the way through the merger.  Unfortunately, there are currently no
methods available for setting up such astrophysically accurate 
initial data.  All existing methods produce BHB initial data with
various spurious effects that deviate somewhat from the desired
astrophysical data.  However, if these deviations are small enough,
they should disappear as the BHs spiral together, leading to 
the correct astrophysical trajectory predicted by the Einstein
equations.

The sequence of runs by the Goddard group clearly demonstrates
this behavior.  Figure~\ref{fig:trajectories}, which is
taken from Ref.~\cite{Baker:2006yw}, shows the
trajectories followed by the punctures in their four runs;
for clarity, only the track of one of the BHs from each simulation
is shown.  Run R4 has the widest initial separation and completes
the most orbits.  After an initial transient period of approximately
one orbit, the trajectory from each of the other runs nearly locks
on to the R4 trajectory.  For the final orbit, the trajectories 
from all of the runs are very nearly superposed.  The fact that
the BH paths lock on to a common universal trajectory for the
final orbit and thereafter supports the idea that the late-time
dynamics is dominated by the strong-field interactions and radiative
losses; this has the effect of reducing the dependence on the
initial conditions.

This universal dynamics produces a universal gravitational
waveform.  Figure~\ref{fig:waves}, which is taken from
Ref.~\cite{Baker:2006yw}, shows one polarization
component of $r\Psi_4$, where $\Psi_4$ is the
Weyl tensor component.  The waveforms have been normalized
so that the peak amplitude of the gravitational radiation
occurs at $t = 0$.  These waveforms all agree to within
$1\%$ for the last orbit, merger and ringdown (after
$t \sim -50M_f$) and, except for a brief initial pulse at
the beginning of each run, to within $\sim 10\%$ for the 
preceeding few orbits (shown in the inset).

The Goddard simulations consistently produce a final merged
remnant BH with spin $a/m = 0.69$ to within $\sim 1\%$.
The amount of energy released as gravitational waves varies slightly,
since the simulations have different durations.  For the longest
run R4, they find $E_{\rm rad} = 0.039M$.

The discussion so far has focused on mergers of nonspinning BHs.
Astrophysical BHs, however, are expected to be spinning.  The UTB
group was the first to carry out merger simulations of equal mass
BHBs with spins \cite{Campanelli:2006uy}. 
Using the moving puncture method, they 
evolved BHs with equal spins, $a = 0.75m$, where $m$ is the 
individual BH mass.  They considered two cases: both BH spins
aligned with the orbital angular momentum, and both spins anti-aligned.
In both cases, the BHs started out on quasi-circular orbits with
period $125M$.

\begin{figure}
 \includegraphics[height=.34\textheight]{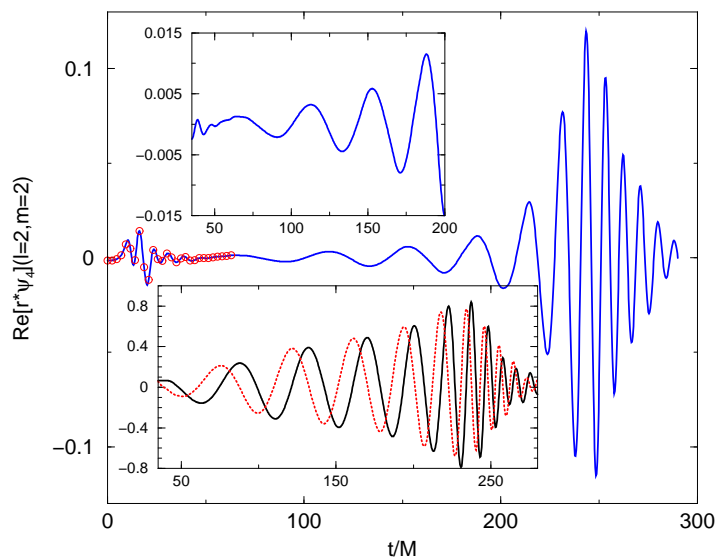}
  \caption{Waveforms from the UTB simulation of merging BHs with spins
aligned with the orbital angular momentum \cite{Campanelli:2006uy}.
The top inset shows the early part of the waveform. The
bottom inset shows the real (solid) and imaginary (dotted)
components of the $l=2, m=2$ component of the strain $h$ calculated
at $r = 10M$.
}
\label{fig:UTBspin}
\end{figure}

The BHB in the anti-aligned case undergoes a prompt merger, completing
$\sim 0.9$ orbits before a common horizon forms, to yield a remnant
BH with spin $a \sim 0.44M$.  In the aligned case, the initial total
angular momentum (orbital + spin) $> M$, and the binary completes
$\sim 2.8$ orbits before merging to form a BH with $a \sim 0.9M$.
In this case, it appears that the merger temporarily stalls as the
excess angular momentum is radiated away, in order to form a final
Kerr BH with $a < M$. 
Figure~\ref{fig:UTBspin}, taken from Ref.~\cite{Campanelli:2006uy}, 
shows the gravitational waveform from this
aligned case.  In both the aligned and anti-aligned cases, 
the resulting gravitational
waveforms show a simple shape, similar to that seen in the non-spinning
case; {\em c.f.} Figure~\ref{fig:waves}

Finally, astrophysical BHBs are also expected to have unequal masses,
especially in the case of the massive BHBs that LISA will observe.
As mentioned above, the gravitational wave emission will be 
asymmetric in this case, imparting a recoil kick to the merged remnant
BH.  Although post-Newtonian techniques have been used to calculate
the kick velocity during the inspiral 
\cite{Blanchet:2005rj,Damour:2006tr}, almost all of the recoil comes
from the strong gravity regime.  Numerical relativity simulations
are therefore needed for an accurate calculation of the kick velocity.
Unequal mass merger calculations are technically more demanding
than the equal mass case, due in part to the need to resolve the
smaller BH which moves faster than the larger one.  In addition,
getting the correct value for the kick requires a sensitive calculation
arising from higher-order gravitational wave modes.

The Penn State group was the first to calculate mergers of unequal mass
nonspinning BHBs, employing 
the moving puncture method \cite{Herrmann:2006ks}.  
Using fairly low resolution and starting their BHs at relatively
close separations, they ran several different mass ratios in the
range $1 \le m_1/m_2 \le 0.32$, and quote lower limits on the kick
velocities.  More recently, the Goddard group carried out more
accurate calculations with higher resolution for the case
$m_1/m_2 = 0.67$ \cite{Baker:2006vn}.  
They also examined the dependence of the
resulting kick velocity on the initial separation of the BHs.
Using higher resolution and adaptive mesh refinement, they
estimate the astrophysically relevant range of kick values
to be $(86 - 97) {\rm km/s}$ for this mass ratio.

\section{Summary and Future Outlook}

The past few years have seen a remarkable series of breakthroughs
in numerical relativity modeling of BHB mergers.  The first BHB
orbit was achieved by Br\"ugmann, Tichy, and Jansen~\cite{Bruegmann:2003aw}
using special comoving gauge conditions in a traditional numerical
relativity code in late 2003.  Roughly a year and a half later, the first
plunge, merger, and ringdown calculation by Pretorius appeared,
using nonstandard techniques and including the extraction of the
gravitational waveform \cite{Pretorius:2005gq}.  

Less than six
months later, in November 2005, the moving puncture method was 
introduced by the UTB \cite{Campanelli:2005dd} and Goddard
\cite{Baker:2005vv} groups, enabling merger calculations with 
simple but novel gauge conditions in traditional numerical
relativity codes.  This was rapidly exploited by its developers
and, by the early spring of 2006, the Goddard group had obtained
a consistent solution for the gravitational wave burst from the merger
of two equal mass Schwarzschild BHs, independent of the 
(quasi-astrophysical) initial conditions \cite{Baker:2006yw}.
Very shortly thereafter, the UTB group produced the first
evolutions of equal mass, spinning BHB mergers, demonstrating the
orbital hangup when the BH spins are aligned with the orbital 
angular momentum \cite{Campanelli:2006uy}.

The first attempt to model unequal mass mergers using the moving puncture
method was carried out by the Penn State group \cite{Herrmann:2006ks}
and first appeared in January 2006.
This was soon followed by similar work at higher resolution and larger
initial separations by the Goddard group \cite{Baker:2006vn}.
Overall, by the summer of 2006, a large segment of the numerical
relativity community was actively adopting, adapting, and exploiting
the moving puncture method.

At the present time there is broad consensus that the merger of
two equal mass Schwarzschild BHs produces a final remnant BH
with spin $a \sim 0.7M$, and that the amount of energy radiated
in the form of gravitatational waves, starting with the final
few orbits and proceeding through the plunge, merger and
ringdown, is $\sim 0.04M$. The UTB and Goddard groups are working with
Pretorius to compare their waveforms; preliminary results using
longer runs by the UTB group and Pretorius show good agreement
with the waveforms obtained by the Goddard group shown in 
Figure \ref{fig:waves}.  Plans are underway to include other
groups in this comparison effort.  

The outlook for continued progress in BHB merger simulations is
very bright.  New work with moving punctures continues to 
appear \cite{Sperhake:2006cy,Hannam:2006vv,Campanelli:2006fg}.
Pretorius is carrying out new and longer runs with his
generalized harmonic code.  The Caltech-Cornell collaboration
has made important progress in carrying out orbits with
their spectral code based on a fully first-order formulation
of the Einstein equations \cite{Scheel:2006gg}
and hopes to achieve mergers soon.  The impressive
progress in this field, occurring across a broad front,
is very encouraging.  Stay tuned!

%%%%%%%%%%%%%%%%%%%%%%%%%%%%%%%%%%%%%%%%%%%%%%%%
%% BACKMATTER
%%%%%%%%%%%%%%%%%%%%%%%%%%%%%%%%%%%%%%%%%%%%%%%%

\begin{theacknowledgments}
  It is a pleasure to acknowledge John Baker,
Jim van Meter, and Richard Matzner
 for valuable comments on the manuscript,
and to thank Frans Pretorius and Manuela Campanelli
for the use of figures from their papers.
This work was supported in part by 
NASA grant O5-BEFS-05-0044.
\end{theacknowledgments}

%%%%%%%%%%%%%%%%%%%%%%%%%%%%%%%%%%%%%%%%%%%%%%%%
%% The bibliography can be prepared using the BibTeX program or
%% manually.
%%
%% The code below assumes that BibTeX is used.  If the bibliography is
%% produced without BibTeX comment out the following lines and see the
%% aipguide.pdf for further information.
%%
%% For your convenience a manually coded example is appended
%% after the \end{document}
%%%%%%%%%%%%%%%%%%%%%%%%%%%%%%%%%%%%%%%%%%%%%%%%

%%%%%%%%%%%%%%%%%%%%%%%%%%%%%%%%%%%%%%%%%%%%%%%%
%% You may have to change the BibTeX style below, depending on your
%% setup or preferences.
%%
%%
%% For The AIP proceedings layouts use either
%%%%%%%%%%%%%%%%%%%%%%%%%%%%%%%%%%%%%%%%%%%%

\bibliographystyle{aipproc}   % if natbib is available
%\bibliographystyle{aipprocl} % if natbib is missing

%%%%%%%%%%%%%%%%%%%%%%%%%%%%%%%%%%%%%%%%%%%
%% You probably want to use your own bibtex database here
%%%%%%%%%%%%%%%%%%%%%%%%%%%%%%%%%%%%%%%%%%%
\bibliography{../bibtex/references.bib}

%%%%%%%%%%%%%%%%%%%%%%%%%%%%%%%%%%%%%%%%%%%
%% Just a reminder that you may have to run bibtex
%% All of it up to \end{document} can be removed
%% if you don't like the warning.
%%%%%%%%%%%%%%%%%%%%%%%%%%%%%%%%%%%%%%%%%%%
\IfFileExists{\jobname.bbl}{}
 {\typeout{}
  \typeout{******************************************}
  \typeout{** Please run "bibtex \jobname" to optain}
  \typeout{** the bibliography and then re-run LaTeX}
  \typeout{** twice to fix the references!}
  \typeout{******************************************}
  \typeout{}
 }

\end{document}